\documentclass[12pt]{article}
\begin{document}
~
~
~
~
\begin{center} {\large \bf  Spectrum of  Particles Created in Inhomogeneous Spacetimes}

\vspace{2cm}
                      Wung-Hong Huang\\
                       Department of Physics\\
                       National Cheng Kung University\\
                       Tainan,\, Taiwan\\
\end{center}
\vspace{2cm}
It is proved that the spectrum of  scalar particles generated from the initial vacuum in inhomogeneous spacetime is nearly thermal in the limit of large momentum $k$, if the momentum was defined as the variable of the Fourier transform of the coordinate in the scalar field. 
\vspace{4cm}
\begin{flushleft}
Physics Letters  B244(1990)378

E-mail:  whhwung@mail.ncku.edu.tw\\
\end{flushleft}
\newpage
Kandrup [ 1 ] has recently proved Parker's conjecture [2] of generating a nearly thermal spectrum from the background geometry of the cosmology. The proof 
could be straightforwardly generalized to anisotropic spacetimes. Let us give a brief sketch. 

Consider a free scalar field $\varphi(x,t)$ in a spatial homogeneous spacetime, which can be viewed as a sum of the modes $\phi_k(x)\,\chi_k(t)$ that evolve independently one from another. Using a suitable time variable (say) $\eta$, e.g., conformal time, and defining a suitable wave function $\Psi_k(\eta) \equiv f (\eta) \chi_k(t)$, where $f(\eta)$ is a function of the metric tensor, then we have the wave equation of  $\Psi_k(\eta)$ (we leave off the index $k$ hereafter) 
$$ {d^2\Psi \over d\eta ^2} + \Omega^2(\eta)\Psi(\eta) = 0. \eqno{(1)}$$ 
In the interval we are interested in $\Omega$ is a real, positive and finite function of the wave vectors $k$, particle mass $m$, curvature coupling constant $\xi$ and metric forms $g_{\mu\nu}$. For example, in the case of the four-dimensional Bianchi type I spacetime with line element 
$$ds^2 = dt^2 - \sum_i\, C_i(t) dx_i^2, \eqno{(2)}$$
then, after the calculation and using the conformal time variable $\eta \equiv \int C(t) dt$, in which $C(t) \equiv (C_lC_2C_3)^{1/3}$,we have [3] 
$$\Omega^2(\eta) = C(\eta) \left( \sum_i{k_i^2\over C_i(\eta)} + m^2 + (\xi -{1\over6})\, R(\eta) \right) + Q(\eta),\eqno{(3)}$$
$$Q(\eta) \equiv {1\over 72}\sum_{i<j} (d_i -d_j)^2,~~~~d_i\equiv {dC_i(\eta)/d\eta\over C_i(\eta)}. \eqno{(4)}$$
The general solution of eq. (1) can be written in the WKB form 
$$\Psi (\eta) = (2\Omega)^{-1/2}\left[\alpha(\eta) \,exp(-i\int^\eta \Omega\, d\eta) + \beta(\eta)\, exp(\int^\eta \Omega\, d\eta)\right]. \eqno{(5)}$$ 
For the two introduced functions $\alpha$ and $\beta$ we can impose one additional condition
$${d\Psi \over d\eta} = - i ({1\over2}\Omega)^{1/2}\left[\alpha(\eta) \,exp(-i\int^\eta \Omega\, d\eta) - \beta(\eta)\, exp(\int^\eta \Omega\, d\eta)\right].\eqno{(6)}$$ 
The Wronskian condition implies that 
$$|\alpha|^2 - |\beta|^2 = 1. \eqno {(7)}$$
Using eqs. (5) and (6) one can reduce the wave equation (1) of  $\Psi$ to two first-order coupled differential equations of $\alpha $ and $\beta$ [4,5]: 
$${d\alpha\over d\eta} = {d\Omega/d\eta\over 2\Omega}\, \beta\, exp\left(2i\,\int^\eta \Omega\, d\eta)\right). \eqno{(8)}$$ 
$${d\beta \over d\eta} = {d\Omega/d\eta\over 2\Omega}\, \alpha\, exp\left(-2i\,\int^\eta \Omega\, d\eta)\right). \eqno{(9)}$$ 
Note that eqs.(5), (6), (8) and (9) are exactly  equivalent to eq.(1), and thus the WKB solution of eq. (5) is an exact form instead of an approximation.

   To make sense of the particle meaning in the initial time ($\eta_1$) and final time ($\eta_2$), the spacetimes in both $\eta_1$ and $\eta_2$ must be either static or adiabatic static [4 ]. Suppose that the Fock spaces at $\eta_1$ and $\eta_2$ are unitarily equivalent, it then follows that $|\beta|^2 \rightarrow 0$ as $k\rightarrow\infty$.  Therefore, in the short-wavelengths limit $\alpha \rightarrow 1$ and
eq. (9) becomes
$${d\beta \over d\eta} = {d\Omega/d\eta\over 2\Omega}\, exp\left(-2i\,\int^\eta \Omega\, d\eta)\right). \eqno{(10)}$$ 
If, for sufficiently large $k$, the $\Omega$ term could be approximated as $k\,\beta(\eta)$ for a suitable function $\beta(\eta)$, then, in a first approximation, we could integrate the above equation and obtain
$$\beta = \int {d\, b/d\eta\over 2b}\, exp\left(-2\,i\,k\,\int^\eta \, b\,d\eta\right)\, d\eta. \eqno{(11)}$$ 

   We can define a new time variable
$$T(\eta) \equiv \int\, b(\eta)\, d\eta ,        \eqno{(12)}$$ 
and then write eq. (11) as
$$\beta = \int\, F(T)\, exp\left(-2\,i\,k\,T\right) \, dT ,$$
$$  F(T) \equiv {d^2 T(\eta)/d\eta^2\over2 [dT(\eta)/d\eta]^2}\eqno{(13)}$$ 
Note that when $b(\eta)$ is a monotonic function (increasing or decreasing) then the function $T$ is also a monotonic function of $\eta$ and thus $T$ may be regarded as a time variable. When $b(\eta)$ is a more general function, we can cut the time variable $\eta$ into several intervals during which $b(\eta)$ is a monotonic function and thus, in each interval, $T$ may still be regarded as a
time variable.

  For a sufficiently well behaving $F(T)$ (this is implied from a sufficiently good behavior of the metric form $g_{\mu\nu}$) and presuming that $T\rightarrow \pm \infty$ for $\eta \rightarrow \eta_{1,2}$, Kandrup then used the method of residues to evaluate the above integration and the thermal distribution of $exp(-const. \,k)$ was obtained.  In this case, however, it is necessary, to assume that $F(t)$ is analytic in the lower-half plane except for isolated poles.

  For the case of $T\not=\pm \infty$ when $ \eta \rightarrow \eta_{1,2}$ we could, in the spirit of Landau's analysis of electrostatic damping [6], argue that for large $|k|$, the integral in eq.(13) should be dominated by an exponentially
 damped function $exp(-const. \,k)$. Thus a nearly thermal form will also show up.

    An obviously physical interpretation for such a  general character is that the short-wavelength modes do not probe the detailed properties of the back-
 ground spacetimes, and thus should evolve in a fairly general fashion [1]. If this interpretation is correct  then one might expect that the thermal spectrum
 should also be generated from an inhomogeneous background spacetime. However, the lack of homogeneity will lead the different mode solutions of the
 curved-space field equation to be mixed, and a key equation like the form of eq. (1) could not be obtained. Therefore, in order to study particle creation
 in inhomogeneous spacetimes one restores the inhomogeneous perturbations of spatial homogeneous cosmologies [7]. As such a method could only treat
 the problems with small inhomogeneities, for studying a general situation we will in this paper directly analyze the mode-mixed equations. It will be found  that, in the limit of large momentum $k$, the mode-mixed equations become mode-separated equations which evolve independently. Also, the general form
of these mode-separated equations are just the form of eq. (1) and thus the thermal behavior arises as well. 

    For convenience, we will only consider the theories in ($1 +1$) dimensions, the reader could easily see that  there is no problem to extend the models to any
dimension. 

    The most general differential equation describing  a scalar field in an inhomogeneous spacetime is
$${d^2\phi\over dt^2} + E(x,t) \,{d^2\phi\over dx^2} + D(x,t) \,{d^2\phi\over dxdt} + C(x,t) \,{d\phi\over dx} + B(x,t) \,{d\phi\over dt}+ A(x,t) \, \phi = 0 , \eqno{(14)}$$ 
where $A(x, t),...E(x, t)$ are functions of the particle mass, curvature coupling constant and metric forms, which need not to be specified in our proof.

   Defining  $\tilde A_k(t), ...,\tilde E_k(t)$, and $\tilde \phi_k(t)$, as the Fourier transforms, with respect to the variable x, of  $A(x, t), ..., E(x, t)$ and $\phi(x, t)$ we can obtain from the above equation

$$\ddot{\tilde \phi}_k - \int_{-\infty}^{\infty}(k-l)^2 \tilde E_l\,\tilde\phi_{k-l}- i  \int_{-\infty}^{\infty}(k-l) (\tilde D_l\,\dot{\tilde\phi}_{k-l}+\tilde C_l\, \tilde\phi_{k-l}) \, dl + \int_{-\infty}^{\infty}(\tilde B_l\,\dot{\tilde\phi}_{k-l}+\tilde A_l\, \tilde\phi_{k-l}) \, dl = 0 , \eqno{(15)}$$ 
\\
where an overdot means $d/dt$. This differential-integral equation manifests mixing between the different modes as a consequence of the inhomogeneity of 
the spacetimes. 

   Using the discussions described below eq. (13) we know that the exponential damping factor $exp( -const.\, l)$ should dominate in the limit of large $l$ for the functions $\tilde A_l(t), ...,\tilde E_l(t),$. Thus, in the first approximation of large $k$, we can reduce the above equation to 
$$\ddot{\tilde \phi}_k - k^2 \tilde E\,\tilde\phi_k- i \,k (\tilde D\,\dot{\tilde\phi}_k+\tilde C\, \tilde\phi_k) + (\tilde B\,\dot{\tilde\phi}_k + \tilde A\,\tilde\phi_k) \, = 0 , \eqno{(16)}$$ 
where 
$$\tilde E \equiv  \int_{-\infty}^{\infty} \tilde E_l\, dl,~~\tilde D \equiv  \int_{-\infty}^{\infty} \tilde D_l\, dl,~~\tilde C \equiv  \int_{-\infty}^{\infty} \tilde C_l\, dl,~~\tilde B \equiv  \int_{-\infty}^{\infty} \tilde B_l\, dl,~~\tilde A \equiv  \int_{-\infty}^{\infty} \tilde A_l\, dl,~~\eqno{(17)}$$ 
One then finds that in the limit of large momentum k the mode-mixed evolution equation becomes a mode-separated evolution equation in any inhomogeneous spacetime. 

Furthermore, as $k$ is very large, we can, in the case of sufficiently well behaving functions $A(t),..., E(t)$, approximate eq. (16) by 
$$\ddot{\tilde \phi}_k - k^2 \tilde E\,\tilde\phi_k- i \,k \,\tilde D\, \dot {\tilde\phi}_k \, = 0 , \eqno{(18)}$$ 
After defining the wave function $\Psi_k(t) \equiv f(t)\, {\tilde\phi}_k $ , in which 
$$ f_k(t) = exp\left(-{1\over2}\, i\, k \int \tilde D (t)\, dt \right), \eqno{(19)}$$
we then obtain the wave equation of $\Psi_k(t)$
$$\ddot\Psi_k(t) + \Omega^2\, \Psi(t) = 0 , ~~~\Omega^2 = k^2\, \left({1\over4}\tilde D^2-\tilde E \right) . \eqno{(20)}$$
As the above equation has the same form as the key equation, eq. (1), we could thus use the same procedure as described in the earlier section to prove that 
the particle spectrum generated from initial vacuum in inhomogeneous spacetime is nearly thermal in the limit of large momentum $k$, if the momentum was defined as the variable of the Fourier transform of the coordinate in the scalar field. 

Note that one needs not worry about the situation of raising a non-positive value of $\Omega ^2(t)$ in the last equation. In fact, only for the cases with reasonable metric forms and after choosing a suitable time variable could one make sense of the particle meaning in the initial and final times, which are necessary conditions for the study of particle creation in curved space. Under such a condition one sees that the particle will propagate with a wave form, and thus the value of the $\Omega ^2(t)$ term in the last equation must always be positive. 

Finally, we want to remark that the scale above which the spectrum is exponential may be determined by the characteristic length over which the universe is inhomogeneous. The spectrum can at best be thermal only for wavelengths short compared to this characteristic length. 
\newpage
\begin{center} {\large \bf  References} \end{center}
\begin{enumerate}
\item  H. E. Kandrup, Phys. Lett. B 215 (1988) 473. 
\item  L. Parker, Nature 261 (1976) 20. 
\item  N. D. Birrell and P. C. W. Davies, "Quantum fields in curved  space", (Cambridge U.P., Cambridge, 1982) Ch. 5.6. 
\item  Ya. B. Zei'dovich and A. A. Starobinsky, Zh. Eksp. Teor. Fiz. 61 (1972) 2162 [Sov.Phys.JETP 34 (1972) 1159]. 
\item  B. L. Hu, Phys. Rev. D 9 (1974) 3263. 
\item  L. D. Landau, Zh. Eksp. Teor. Fiz. 16 (1946) 574. 
\item  J. A. Frieman, Phys. Rev. D 39 (1989) 389. 
\end{enumerate}
\end{document}